\documentclass[aps,prd,amsmath,amssymb,showpacs]{revtex4}

\usepackage{graphicx}
\begin{document}

\title{Indirect search for lepton flavour violation 
at CERN LEP via  doubly charged bileptons }
\author{S. Ata\u{g}}
\email[]{atag@science.ankara.edu.tr}
\affiliation{Department of Physics, Faculty of Sciences,
Ankara University, 06100 Tandogan, Ankara, Turkey}

\author{K.O. Ozansoy }
\email[]{oozansoy@science.ankara.edu.tr}
\affiliation{Department of Physics, Faculty of Sciences,
Ankara University, 06100 Tandogan, Ankara, Turkey}

\begin{abstract}
We search for lepton flavour violating couplings via 
doubly charged bilepton (or doubly charged Higgs) exchange 
in electron-positron annihilation
process $ \text{e}^{+}\text{e}^{-}\to \mu^{+}\mu^{-},
\tau^{+}\tau^{-}$ using CERN LEP data at the center of 
mass energies between 189-207 GeV.
Standard model program ZFITTER
has been used to calculate radiative corrections. We find that 
$g_{L}^{2}/M_{L}^{2}<O(10^{-6})GeV^{-2}$ at 95\% C.L. 
for flavour violating scalar and vector bilepton couplings.
\end{abstract}


\maketitle

\section{Introduction}
In the minimal Standard Model (SM) the usual Higgs mechanism 
responsible for the electroweak symmetry breaking implies the 
conservation of the lepton number separately for each generation.
As is well known, the current low energy phenomenology of the SM 
is quite consistent with all present experiments. However, there 
has been no experimental evidence for the existence of the SM 
Higgs boson. This is one of the good reasons that other symmetry 
breaking mechanisms and extended Higgs sectors have not been 
excluded in the theoretical point of view. In addition, 
indication for neutrino oscillations necessarily violate 
lepton-flavour symmetry\cite{fukuda}.
In the theories beyond the SM, doubly charged, lepton flavour 
changing, exotic bosons may occur. Models with extended 
Higgs sectors include doubly charged Higgs boson \cite{haber}.
The supersymmetric extensions, such as SO(10) SUSY GUT model 
where supersymmetric lepton partners induce lepton 
flavour violation \cite{barbieri}.
The purpose of this paper is to study lepton flavour 
violation through the possible existence of doubly 
charged bileptons.
Bileptons  are defined as bosons carrying lepton 
number L=2 or 0 which  couple to two standard model 
leptons but not to quarks. Bileptons appear in models where 
$\text{SU(2)}_{\text{L}}$ gauge group is extended 
to SU(3) \cite{frampton} and   
also  in models with extended Higgs sectors that contain 
doubly charged Higgs bosons \cite{haber}. Grand unified theories, 
technicolor and composite models predict 
the existence of bileptons as well as 
other exotic particles \cite{ross}. Classification and interactions of 
bileptons are provided by several works \cite{rizzo} and  
a comprehensive review has been presented  in \cite{cuypers}
including low and high energy bounds on their couplings.
Indirect constraints on the masses and couplings of doubly 
charged bileptons have been obtained from $\mu$ and $\tau$ 
physics,  muonium-antimuonium conversion and Bhabha scattering 
experiments \cite{swartz,sasaki,chang,willmann,okan}. 
Searches for doubly charged Higgs  
have been performed by DELPHI and OPAL collaborations at LEP
\cite{delphi, opal}.

At LEP, fermion pair production is the unique 
reaction to test the standard model at loop
correction level \cite{sirlin}. 
Therefore one needs precision calculations 
including QED and weak corrections for reliable comparison 
with experiments. ZFITTER is one of the standard model 
program developed to compute scattering cross sections 
and asymmetries for fermion pair production 
in $\text{e}^{+}\text{e}^{-}$ 
collision with QED and electroweak corrections
\cite{zfitter}. Using cross section
calculations with ZFITTER realistic limits for new physics 
can be obtained from LEP data. 
Since QED corrections are model independent (well-defined 
if couplings, masses and widths of new particles are fixed), the usual 
convolution formulae can be applied for total cross section 

\begin{eqnarray}
\sigma(s)=\int{dv~ \sigma^{0}(s^{\prime})~R(v)} 
\end{eqnarray}
with $v=1-s^{\prime}/s$. For this reason, the flux factor $R(v)$ 
(radiator) is not influenced by new particles.
Above equation can be straightforwardly generalized to 
different asymmetries $A_{FB}$, $A_{pol}$, $A_{LR}$ or to 
scattering angle distribution $d\sigma/\cos{\theta}$ 
with different effective Born terms and radiators. Final state 
acollinearity cut and minimum energy can also be applied.
Contribution of doubly charged bileptons to 
 ${e}^{+}{e}^{-}$ annihilation process 
is the subject of the present article  

\begin{eqnarray}
\text{e}^{+}\text{e}^{-}\to (\gamma, Z, L^{--})\to
\mu^{+}\mu^{-} (\gamma), ~~
\tau^{+}\tau^{-} (\gamma)
\end{eqnarray}
where $L^{--}$ is the doubly charged bilepton giving flavour violating 
contribution. Here ($\gamma$) represents initial
and final state radiations. 

General effective lagrangian describing interactions of 
bileptons with the standard model leptons is generated by 
requiring $\text{SU(2)}_{\text{L}}\times \text{U(1)}_
{\text{Y}}$ invariance. We consider  the lagrangian involving 
the bilepton couplings to leptons only for L=2 bileptons as follows: 

\begin{eqnarray}
{\cal L}_{L=2}=&&g_1^{ij} \bar{\ell}_{i}^{c}i\sigma_2 
\ell_{j} L_1+\tilde{g}_1^{ij}
\bar{e}_{iR}^{c}e_{jR}\tilde{L}_1 \nonumber \\
&&+g_2^{ij}\bar{\ell}_{i}^{c}i\sigma_2\gamma_{\mu}
e_{jR}L_{2}^{\mu}+g_3^{ij}\bar{\ell}_{i}^{c}i\sigma_2\vec{\sigma}
\ell_{j}~.\vec{L}_3+h.c.
\end{eqnarray}

In the notations we have used ${\ell}$ is the left handed 
$\text{SU(2)}_\text{L}$ lepton doublet and $e_{R}$ is the 
right handed charged singlet lepton. Charge conjugate fields are 
defined as $\psi^{c}=C\bar{\psi}^T$ and $\sigma_1$, 
$\sigma_2$, $\sigma_3$ are the Pauli matrices.  
The subscript of bilepton fields $L_{1,2,3}$ and 
couplings $g_{1,2,3}$ denote $\text{SU(2)}_\text{L}$ singlets, 
doublets and triplets. We show flavour indices by $i,j=1,2,3$.
Here we are interested only in doubly charged bileptons. 
In order to express the lagrangian in terms of individual electron, 
bileptons and helicity projection operators 
$\text{P}_{\text{R/L}}=\frac{1}{2}(1\pm\gamma_5)$
we expand the Pauli matrices and lepton doublets.
Then we write the lagrangian as :

\begin{eqnarray}
{\cal L}_{L=2}=&&\tilde{g}_1~\tilde{L}_1^{++}~\bar{e}^{c}P_R e
+g_2~ L_{2\mu}^{++}~\bar{e}^{c}\gamma^{\mu}P_L e
\nonumber\\
&&-\sqrt{2}g_3~ L_3^{++}~\bar{e}^{c}P_L e + h.c.
\end{eqnarray}
where superscripts of bileptons stand for their electric charges.
For simplicity, flavour indices have been skipped. 
If the scalar $L_3^{--}$ acquires a vacuum expectation value then 
it is a doubly charged Higgs boson that appears in 
the left-right symmetric models \cite{rizzo}.

\section{Electron-Positron Annihilation Processes
$e^{+}e^{-}\to \mu^{+}\mu^{-}$, $\tau^{+}\tau^{-}$ }

Including the doubly charged scalar bilepton 
 $\text{L}_{3}^{--}$ exchange
with lepton flavour changing couplings,
the unpolarized differential cross section for the proceess 
$e^{+}e^{-}\to \ell^{+}\ell^{-}$ $(\ell=\mu, \tau)$ 
in terms of mandelstam invariants s, t and u is given by

\begin{eqnarray}
\frac{d\sigma}{d\cos{\theta}}=&&\frac{\pi\alpha^{2}}{8s}
\left\{ \left[ 2\frac{u}{s}+2C_L^{2}\frac{u}{s-M_{Z}^2}-
\frac{g_L^2}{g_e^2}\frac{u}{u-M_{L}^2}\right]^2 \right. \nonumber\\
&&\left.+\left[2\frac{u}{s}+2C_R^{2}\frac{u}{s-M_{Z}^2}\right]^2 
+2\left[\frac{2t}{s}+C_L C_R\frac{2t}{s-M_{Z}^2}\right]^2 \right\}.
\end{eqnarray}

With the flavour violating  doubly charged vector bilepton couplings
($\text{L}_{2\mu}^{--}$ exchange with coupling $g_2$),
the unpolarized differential cross section for the above process 
takes the form  

\begin{eqnarray}
\frac{d\sigma}{d\cos{\theta}}=&&\frac{\pi\alpha^{2}}{8s}
\left\{ \left[ 2\frac{u}{s}+2C_L^{2}
\frac{u}{s-M_{Z}^2}\right]^2 \right.  \nonumber \\
&&\left.+\left[2\frac{u}{s}+2C_R^{2}\frac{u}{s-M_{Z}^2}\right]^2
+\left[\frac{2t}{s}+C_LC_R\frac{2t}{s-M_{Z}^2}\right]^2 \right. 
\nonumber \\
&&\left.+\left[\frac{2t}{s}+C_LC_R\frac{2t}{s-M_{Z}^2}+
\frac{g_L^2}{g_e^2}\frac{2t}{u-M_{L}^2}\right]^2
\right\}
\end{eqnarray}
where the definition of mandelstam variables and
$C_L$,  $C_R$ are as follows:
\begin{eqnarray}
&&t=-\frac{s}{2}(1-\cos{\theta}), ~~
u=-\frac{s}{2}(1+\cos{\theta}),  \\
&&C_L=\frac{2\sin^2{\theta_W}-1}{2\sin{\theta_W}\cos{\theta_W}},
~~C_R=\tan{\theta_W}.
\end{eqnarray}
Bilepton couplings $\sqrt{2}g_{3}$ and $g_2$ in the lagrangian
have been replaced by $g_{L}$ in the cross sections.
Electromagnetic coupling $g_e$ is defined by $g_{e}^{2}=4\pi\alpha$.

We use the conventional  approach to correcting the process
$e^{+}e^{-} \to X \to f\bar{f} $ for radiative effects
where $X$ represents an exchanged boson arising from
new physics.
For the new physics predictions of an observable 
$O^{\prime}_{NP}$ with  radiative effects,  
the tree level new physics predictions of an 
observable $O_{NP}$  is multiplied by a factor of the SM prediction 
including radiative corrections $O^{\prime}_{SM}$ divided by 
the SM prediction at tree level $O_{SM}$,

\begin{eqnarray}
O^{\prime}_{NP}=O_{NP} \times \frac{O^{\prime}_{SM}}{O_{SM}}
\end{eqnarray}

\begin{figure}
\includegraphics{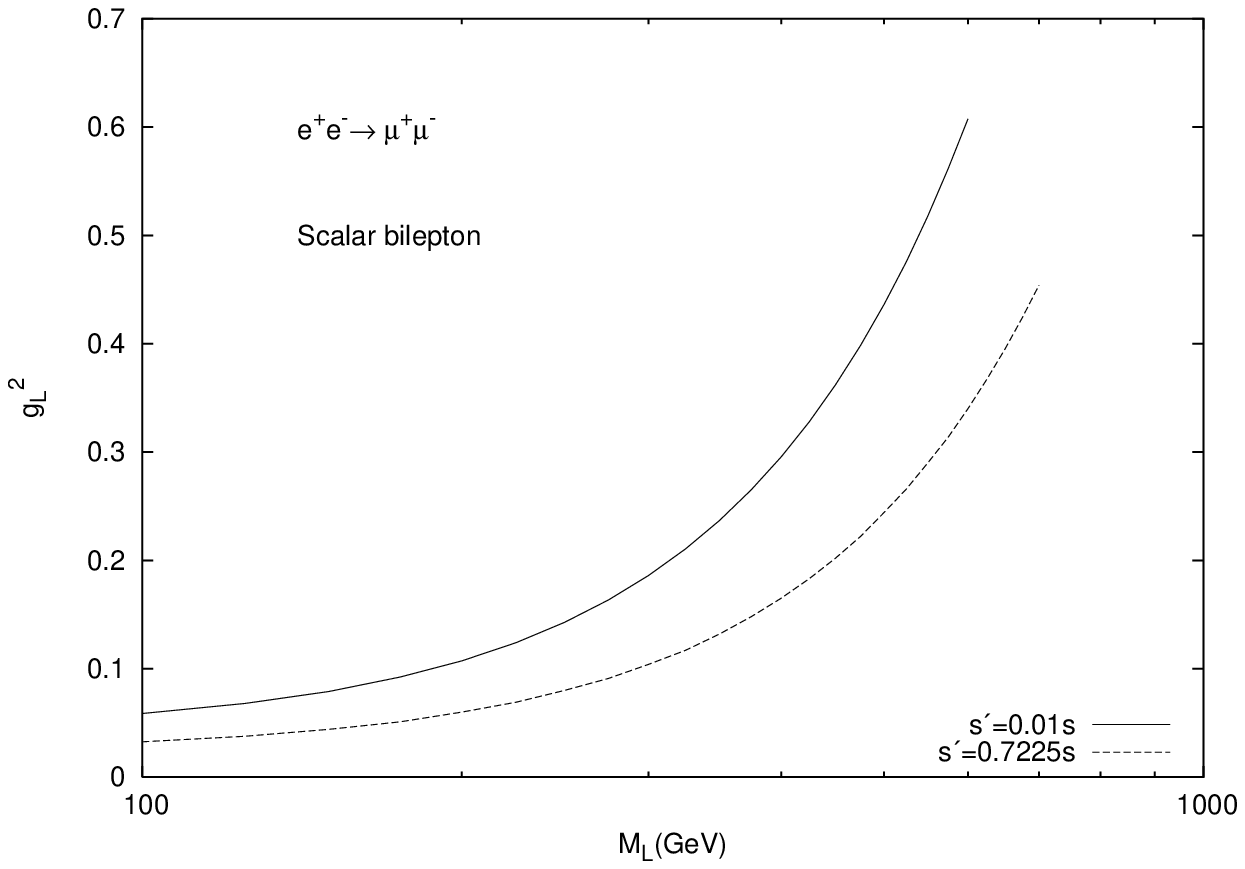}
\caption{Upper limits on the scalar bilepton couplings
squared against bilepton masses obtained from
$e^{+}e^{-}\to \mu^{+}\mu^{-}$ process. \label{fig1}}
\end{figure}

\begin{figure}
\includegraphics{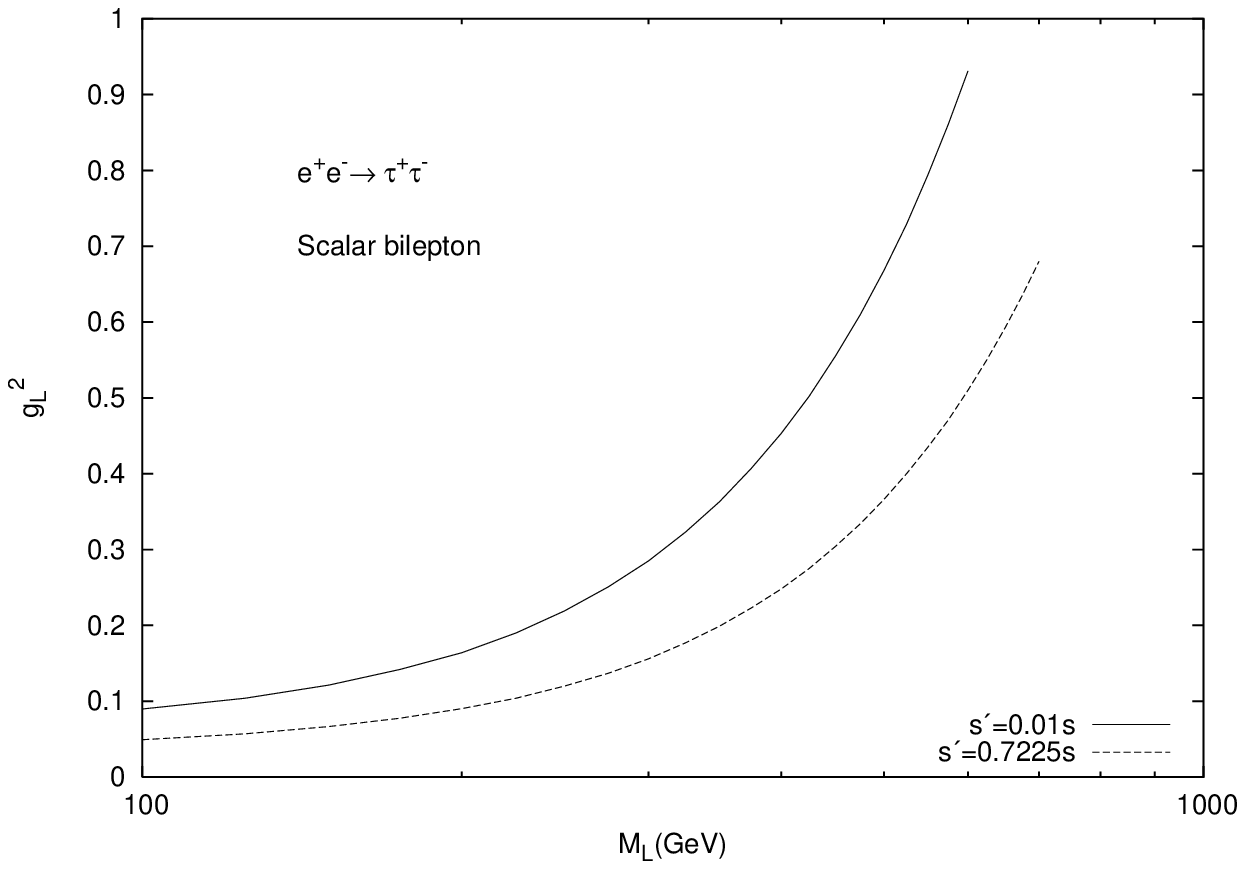}
\caption{Upper limits on the scalar bilepton couplings
squared against bilepton masses obtained from
$e^{+}e^{-}\to \tau^{+}\tau^{-}$ process.
\label{fig2}}
\end{figure}

\begin{figure}
\includegraphics{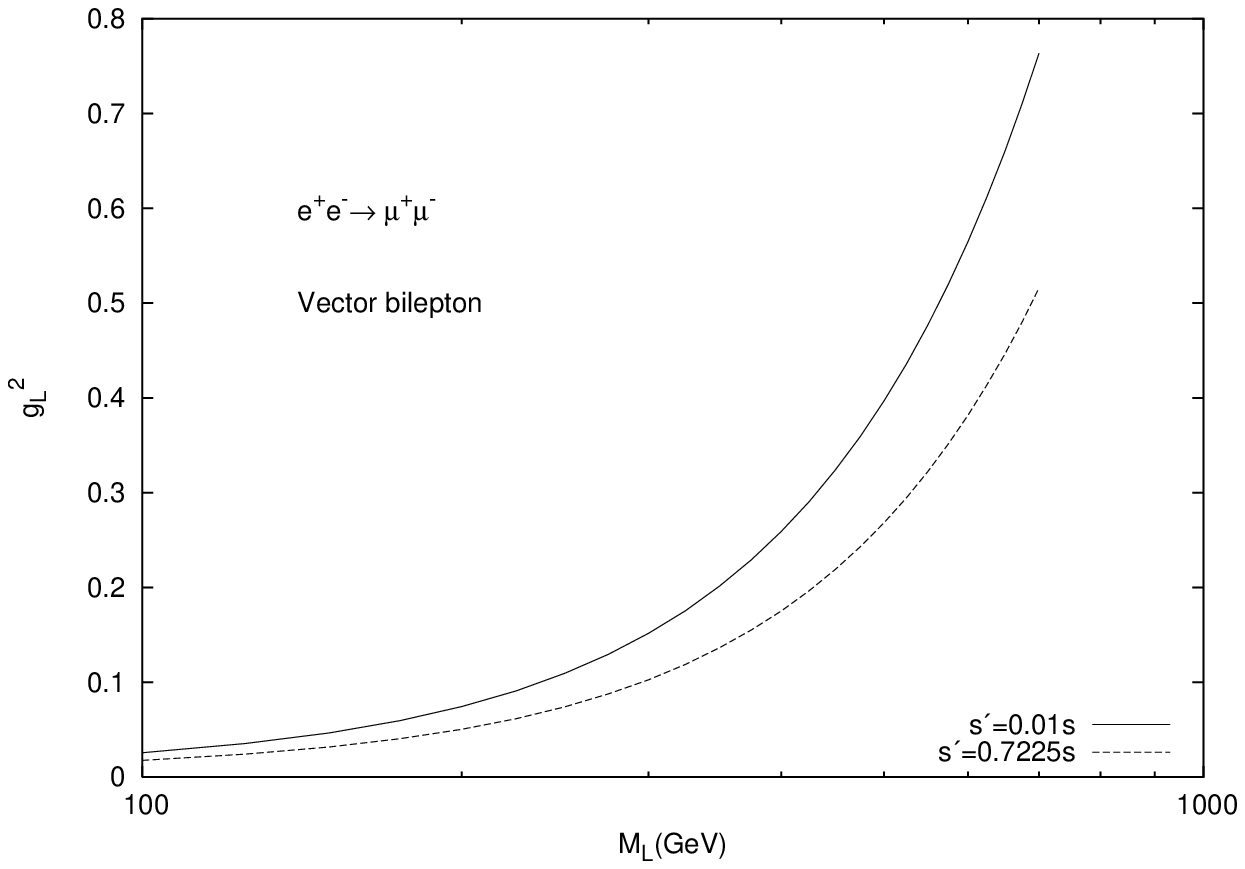}
\caption{Upper limits on the vector bilepton couplings
squared against bilepton masses obtained from
$e^{+}e^{-}\to \mu^{+}\mu^{-}$ process.
\label{fig3}}
\end{figure}

\begin{figure}
\includegraphics{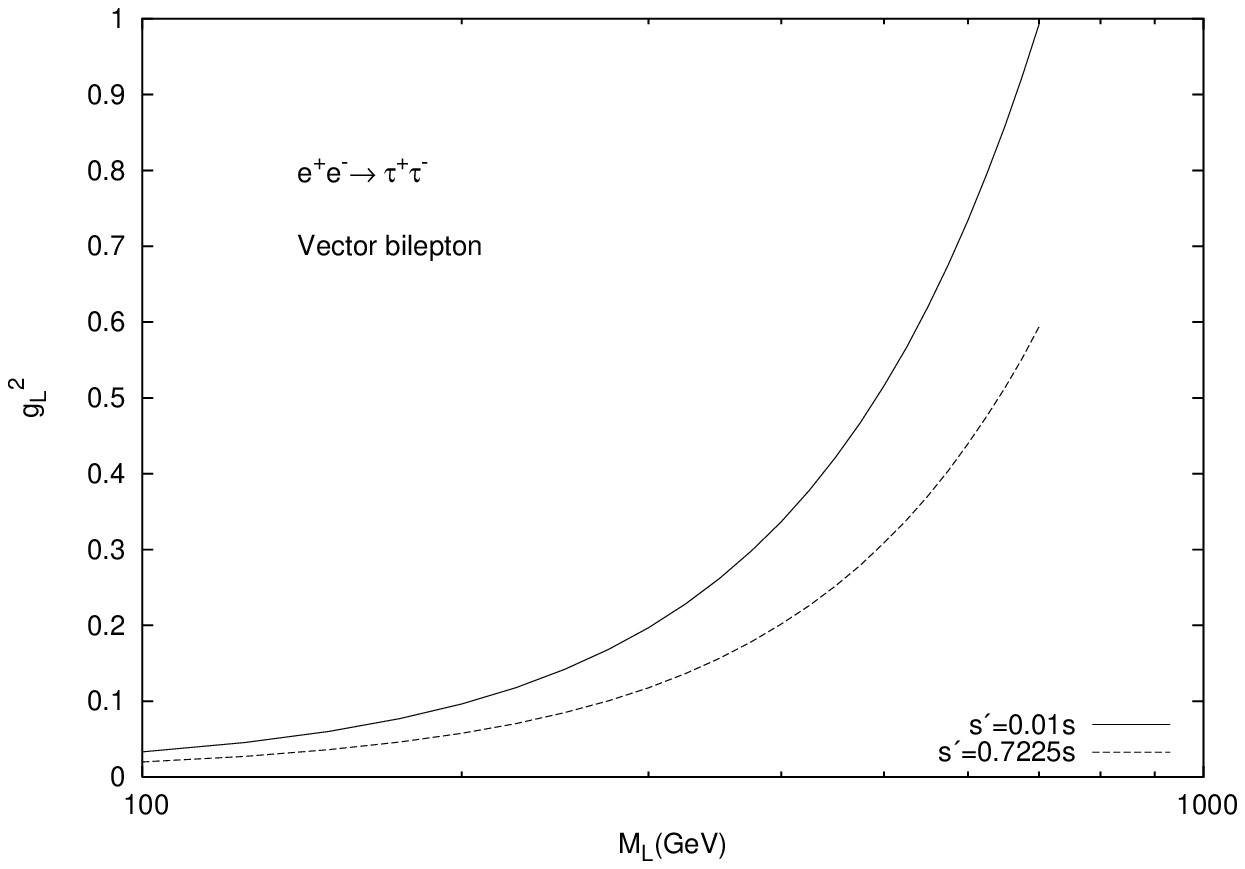}
\caption{Upper limits on the vector bilepton couplings
squared against bilepton masses obtained from
$e^{+}e^{-}\to \tau^{+}\tau^{-}$ process.
\label{fig4}}
\end{figure}

In order to estimate upper limits on $g_{L}$, one parameter, 
one sided $\chi^{2}$ analysis has been used by varying the 
$g_{L}^2$ for fixed bilepton mass between 100 GeV and 700 GeV. 
After 700 GeV bilepton coupling $g_{L}$  exceeds perturbative 
region, approaching unity. 
Total cross sections measured by OPAL detector at CERN LEP for
$\mu^{+}\mu^{-}$ and  $\tau^{+}\tau^{-}$ final states
are given in Table~\ref{tab1} and Table~\ref{tab2} with
standard model values predicted by ZFITTER at energy region
$\sqrt{s}=189-207$ GeV \cite{opal2}. The cross sections and 
errors shown on these tables are used in the following $\chi^{2}$
expressions 

\begin{eqnarray}
\chi^{2}=\sum_{i}(\frac{\sigma_{i}^{exp}-\sigma_{i}^{new}}
{\Delta_{i}^{exp}})^{2}
\end{eqnarray}

\begin{eqnarray}
\Delta^{exp}=\sigma^{exp}\sqrt{\delta_{stat}^{2}+
\delta_{sys}^{2}}
\end{eqnarray}
where $i$ represents energy index corresponding energy values,
cross sections and experimental errors in Table~\ref{tab1} or 
Table~\ref{tab2}.
Fig.~\ref{fig1} and Fig.~\ref{fig2} show the upper  limits 
at 95\% confidence level  on the square of the scalar bilepton 
coupling  $g_{L}^2$ for $e \to \mu$  and $e \to \tau$ type 
lepton flavour violation as a function of bilepton mass.
Upper limits on flavour changing 
vector bilepton couplings squared are shown 
in Fig.~\ref{fig3} and Fig.~\ref{fig4}. 
For higher bilepton masses than LEP energies 
$s<<M_{L}^{2}$ 
we take into account the approximation
in the bilepton propagator:

\begin{eqnarray}
\frac{g_{L}^2}{u-M_{L}^2}\simeq-\frac{g_{L}^2}{M_{L}^2}
\end{eqnarray}
which reduces the number of parameters.
Using the similar $\chi^{2}$ methods we obtain upper 
limits on the $g_{L}^2/M_{L}^{2}$  shown in 
Tables~\ref{tab3}-~\ref{tab6} for the same  kind of 
flavour violating cases  as in the figures. 
From tables combined upper limit on $g_{L}^2/M_{L}^{2}$ is 
about $1\times 10^{-6}$. We use the same ZFITTER version 
and the parameters as the ones used by the OPAL 
Collaboration in Ref.~\onlinecite{opal2} for consistency 
in radiative corrections.

In conclusion previous limits in units of GeV$^{-2}$ 
from muon and tau physics for 
flavour violating doubly charged  
bilepton couplings are given below 
to compare with our results: 
\begin{eqnarray}
g_{ee}g_{e\mu}/M_{L}^{2}<4.7\times 10^{-11} 
~~\,\,\,\,\,\, \mu \to ee\bar{e} \\ 
g_{fe}g_{f\mu}/M_{L}^{2}<2\times 10^{-9} 
~~\,\,\,\,\,\,\mu \to e\gamma \\
g_{f\mu}g_{f\mu}/M_{L}^{2}<4\times 10^{-4} ~~ \,\,\,\,\,\,
(g-2)_{\mu} \\
g_{\tau\mu}g_{e\mu}/M_{L}^{2}<4\times 10^{-7} ~~\,\,\,\,\,\,
\tau \to \ell \ell \bar{\ell} \\
g_{f\tau}g_{fe}/M_{L}^{2}<4\times 10^{-6} ~~\,\,\,\,\,\,
\tau \to e\gamma \\
g_{f\tau}g_{f\mu}/M_{L}^{2}<7\times 10^{-6} ~~\,\,\,\,\,\,
\tau \to \mu\gamma 
\end{eqnarray}

where the first constraint is from Ref.~\onlinecite{swartz}
at 90\% C.L. and others are from Ref.~\onlinecite{cuypers} at 
$2\sigma$. In $\mu$ and $\tau$ decays, the labels $\ell$ stand
for $e$ or $\mu$ whereas the labels $f$ stand for all three 
lepton families.

Concerning the LEP experiments and new physics, two comments are 
in order.
At LEP1 and LEP2 energies the forward-backward asymmetries
have been accurately measured for two fermion final
states\cite{opal2,opal3} . These measurements  can also be
used  for the analysis of the new physics effect.

The upper limits on the  cross sections
for directly flavour violating events 
$e^{+}e^{-}\to e\mu, e\tau, \mu\tau$ were obtained at LEP2 
energies \cite{opal4}. The limits which were obtained range
from 22 fb to 58 fb for $e\mu$ channel, from 78 fb to 144 fb 
for $e\tau$ channel. Such  small cross sections should have 
direct impact on the natural values of flavour 
violating couplings.

\newpage

\newpage

\begin{table}
\caption{Measured cross sections for
$e^{+}e^{-}\to\mu^{+}\mu^{-}$ with OPAL detector
at LEP for two different $s^{\prime}/s$ cuts.
The  first error shown is statistical, 
the second systematic. The standard model
cross sections are predicted by ZFITTER\label{tab1}}
\begin{ruledtabular}
\begin{tabular}{llrlr}
$\sqrt{s}$ GeV & $\sigma$(pb) & $\sigma^{SM}$(pb) &$\sigma$(pb) &
$\sigma^{SM}$(pb)\\
&$s^{\prime}/s> 0.01$ &$ s^{\prime}/s>0.01$ &$s^{\prime}/s> 0.7225$ &
$s^{\prime}/s> 0.7225$\\
\hline
189 & 7.85$\pm$0.25$\pm$0.09 &7.75 & 3.14$\pm$0.15$\pm$0.03 &3.21 \\
192 &7.40$\pm$0.61$\pm$0.09 & 7.47 & 2.86$\pm$0.34$\pm$0.03 &3.10 \\
196 &7.08$\pm$0.37$\pm$0.08 & 7.13 &2.93$\pm$0.22$\pm$0.03 & 2.96\\
200 &6.67$\pm$0.36$\pm$0.08 & 6.80 &2.77$\pm$0.21$\pm$0.03 & 2.83 \\
202 &5.63$\pm$0.48$\pm$0.07 & 6.64 &2.36$\pm$0.28$\pm$0.03 &2.77 \\
205 & 6.53$\pm$0.35$\pm$0.08& 6.41 & 2.88$\pm$0.21$\pm$0.03 &2.67 \\
207 & 6.81$\pm$0.28$\pm$0.08& 6.29 & 2.77$\pm$0.16$\pm$0.03 &2.63 \\
\end{tabular}
\end{ruledtabular}
\end{table}
  
\begin{table}
\caption{Measured cross sections for
$e^{+}e^{-}\to\tau^{+}\tau^{-}$ with OPAL detector
at LEP for two different $s^{\prime}/s$ cuts.
The  first error shown is statistical,
the second systematic. The standard model
cross sections are predicted by ZFITTER\label{tab2}}
\begin{ruledtabular}
\begin{tabular}{llrlr}
$\sqrt{s}$ GeV & $\sigma$(pb) & $\sigma^{SM}$(pb) &$\sigma$(pb) &
$\sigma^{SM}$(pb)\\
&$s^{\prime}/s> 0.01$ &$ s^{\prime}/s>0.01$ &$s^{\prime}/s> 0.7225$ &
$s^{\prime}/s> 0.7225$\\
\hline
189 & 8.17$\pm$0.39$\pm$0.21 &7.74 & 3.45$\pm$0.21$\pm$0.09 &3.21 \\
192 &7.74$\pm$0.95$\pm$0.20 & 7.47 & 3.17$\pm$0.50$\pm$0.08 &3.10 \\
196 &7.21$\pm$0.57$\pm$0.19 & 7.12 &2.89$\pm$0.30$\pm$0.07 & 2.96\\
200 &7.04$\pm$0.56$\pm$0.18 & 6.80 &3.14$\pm$0.30$\pm$0.08 & 2.83 \\
202 &7.69$\pm$0.84$\pm$0.20 & 6.63 &2.95$\pm$0.43$\pm$0.07 &2.77 \\
205 & 6.84$\pm$0.55$\pm$0.18& 6.40 & 2.72$\pm$0.28$\pm$0.07 &2.67 \\
207 & 6.39$\pm$0.41$\pm$0.17& 6.28 & 2.78$\pm$0.22$\pm$0.07 &2.63 \\
\end{tabular}
\end{ruledtabular}
\end{table}

\begin{table}
\caption{Upper limits on the $e \to \mu$ type lepton flavor
violating $g_{L}^{2}/M_{L}^{2}$ for
doubly charged scalar bileptons at 95\% C.L. Combination of
results are also shown in the last row  and masses
are in units of GeV. \label{tab3}}
\begin{ruledtabular}
\begin{tabular}{lcc}
$\sqrt{s}$ GeV & ${g_L^2}\over{m_L^2}$ & ${g_L^2}\over{m_L^2}$ \\
 &$s^{\prime}/s>0.01$ &$s^{\prime}/s>0.7225$ \\
\hline
189 &$<3.2\times10^{-6}$   &$<1.7\times10^{-6}$ \\
192 &$<6.1\times10^{-6}$ &$<3.1\times10^{-6}$ \\
196 &$<3.9\times10^{-6}$ &$<2.4\times10^{-6}$ \\
200 &$<3.6\times10^{-6}$ &$<2.2\times10^{-6}$ \\
202 &$<3.3\times10^{-6}$ &$<1.8\times10^{-6}$ \\
205 &$<3.5\times10^{-6}$ &$<2.2\times10^{-6}$ \\
207 &$<2.9\times10^{-6}$ &$<1.7\times10^{-6}$ \\
\hline
Combination& $<1.6\times10^{-6}$ &$<8.7\times10^{-7}$ \\
\end{tabular}
\end{ruledtabular}
\end{table}

\begin{table}
\caption{Upper limits on the $e \to \tau$ type lepton flavor
violating $g_{L}^{2}/M_{L}^{2}$ for
doubly charged scalar bileptons at 95\% C.L. Combination of
results are also shown in the last row  and masses
are in units of GeV. \label{tab4}}
\begin{ruledtabular}
\begin{tabular}{lcc}
$\sqrt{s}$ GeV & ${g_L^2}\over{m_L^2}$ & ${g_L^2}\over{m_L^2}$ \\
&$s^{\prime}/s>0.01$ &$s^{\prime}/s>0.7225$ \\
\hline
189 &$<4.9\times10^{-6}$   &$<2.8\times10^{-6}$ \\
192 &$<8.8\times10^{-6}$ &$<5.3\times10^{-6}$ \\
196 &$<5.8\times10^{-6}$ &$<3.1\times10^{-6}$ \\
200 &$<5.5\times10^{-6}$ &$<3.31.1\times10^{-6}$ \\
202 &$<7.5\times10^{-6}$ &$<4.2\times10^{-6}$ \\
205 &$<5.2\times10^{-6}$ &$<2.9\times10^{-6}$ \\
207 &$<4.1\times10^{-6}$ &$<2.3\times10^{-6}$ \\
\hline
Combination& $<2.4\times10^{-6}$ &$<1.3\times10^{-6}$ \\
\end{tabular}
\end{ruledtabular}
\end{table}

\begin{table}
\caption{Upper limits on the $e \to \mu$ type lepton flavor
violating $g_{L}^{2}/M_{L}^{2}$ for
doubly charged vector bileptons at 95\% C.L. Combination of
results are also shown in the last row  and masses
are in units of GeV. \label{tab5}}
\begin{ruledtabular}
\begin{tabular}{lcc}
$\sqrt{s}$ GeV & ${g_L^2}\over{m_L^2}$ & ${g_L^2}\over{m_L^2}$ \\
&$s^{\prime}/s>0.01$ &$s^{\prime}/s>0.7225$ \\
\hline
189 &$<2.7\times10^{-6}$   &$<2.1\times10^{-6}$ \\
192 &$<7.4\times10^{-6}$ &$<4.2\times10^{-6}$ \\
196 &$<6.0\times10^{-6}$ &$<2.3\times10^{-6}$ \\
200 &$<5.5\times10^{-6}$ &$<2.2\times10^{-6}$ \\
202 &$<4.1\times10^{-6}$ &$<3.0\times10^{-6}$ \\
205 &$<2.8\times10^{-6}$ &$<1.8\times10^{-6}$ \\
207 &$<2.2\times10^{-6}$ &$<1.4\times10^{-6}$ \\
\hline
Combination& $<1.5\times10^{-6}$ &$<1.0\times10^{-6}$ \\
\end{tabular}
\end{ruledtabular}
\end{table}

\begin{table}
\caption{Upper limits on the $e \to \tau$ type lepton flavor
violating $g_{L}^{2}/M_{L}^{2}$ for
doubly charged vector bileptons at 95\% C.L. Combination of
results are also shown in the last row  and masses
are in units of GeV. \label{tab6}}
\begin{ruledtabular}
\begin{tabular}{lcc}
$\sqrt{s}$ GeV & ${g_L^2}\over{m_L^2}$ & ${g_L^2}\over{m_L^2}$ \\
&$s^{\prime}/s>0.01$ &$s^{\prime}/s>0.7225$ \\
\hline
189 &$<3.7\times10^{-6}$   &$<2.2\times10^{-6}$ \\
192 &$<5.9\times10^{-6}$ &$<4.0\times10^{-6}$ \\
196 &$<4.3\times10^{-6}$ &$<5.2\times10^{-6}$ \\
200 &$<4.0\times10^{-6}$ &$<2.4\times10^{-6}$ \\
202 &$<4.7\times10^{-6}$ &$<3.1\times10^{-6}$ \\
205 &$<3.6\times10^{-6}$ &$<2.4\times10^{-6}$ \\
207 &$<3.1\times10^{-6}$ &$<1.9\times10^{-6}$ \\
\hline
Combination& $<2.0\times10^{-6}$ &$<1.2\times10^{-6}$ \\
\end{tabular}
\end{ruledtabular}
\end{table}


\begin{thebibliography}{99}
\bibitem{fukuda}Y. Fukuda et al.,Super-Kamiokande Collaboration,
Phys. Lett. {\bf B 433}(1998)9; 
Phys. Lett. {\bf B 436}(1998)33; 
Phys. Rev. Lett. {\bf 81}(1998)1562;
J.N.Bahcall et al., Nucl. Phys. Proc. Suppl.{\bf 100}(2001)5;
SNO Collaboration, Q.R. Ahmad {\it et al.}, Phys. Rev. Lett. 
87 (2001)071301.
\bibitem{haber} T.G. Rizzo, Phys. Rev. {\bf D 25}(1982)1355;
J.F. Gunion, H.E. Haber, G. Kane, S. Dawson
The Higgs Hunters Guide, Addison-Wesley (1990); K.Huitu
and J. Maalampi, Phys. Lett. {\bf B 344} (1995)217.
\bibitem{barbieri} R. Barbieri, L. Hall and A. Strumi,
Nucl. Phys. {\bf B 445}(1995)219;

Y. Okada, Proceedings of IV International Workshop Held at 
National Sun Yat-Sen University, June 18-21, 1998,
KEK-TH-606, hep-ph/9811502.
\bibitem{frampton} P.H. Frampton, Int. J. Mod. Phys.
{\bf A 13} (1998)2345.
\bibitem{ross} G.G. Ross, Grand Unified Theories,
Benjamin-Cummins (1985); E. Fahri, L. Suskind Phys.
Rep. {\bf 74} (1981) 277; E. Eichten {\it et al.},
Rev. Mod. Phys. {\bf 56} (1984)579; W. Buchm\"{u}ller,
Acta Phys. Austr. Suppl XXVII (1985)517.
\bibitem{rizzo} T. Rizzo, Phys. Rev.  {\bf D 27} (1983)657;
N. Lepore, B. Thorndyke, H. Nadeau and D. London,
Phys. Rev. {\bf D 50}(1994)2031; G. Barenboim, K. Huitu,
J. Maalampi and M. Raidal, Phys. Lett. {\bf B 394} (1997)132;
J.F. Gunion, Int. J. Mod. Phys. {\bf A 13} (1998)2277.
\bibitem{cuypers} F. Cuypers and S. Davidson, Eur. Phys.
Jour. {\bf C 2} (1998)503.
\bibitem{swartz} M.L. Swartz, Phys. Rev. {\bf D 40} (1989)1521.
\bibitem{sasaki} H. Fujii, Y. Mimura, K. Sasaki and T. Sasaki,
Phys. Rev. {\bf D 49} (1994)559.
\bibitem{chang} D. Chang and W. -Y. Keung, Phys. Rev. Lett.
{\bf 62} (1989)2583.
\bibitem{willmann} L. Willmann {\it et al.}, Phys. Rev. Lett.
{\bf 82} (1999) 49.
\bibitem{okan} S. Atag and K.O. Ozansoy, Phys. Rev.{\bf D 68}
(2003) 093008.
\bibitem{delphi} DELPHI Collaboration, J. Abdallah {\it et al.},
Phys. Lett {\bf B 552} (2003)127.
\bibitem{opal} OPAL Collaboration, G. Abbiendi {\it et al.},
Phys. Lett. {\bf B 577 } (2003)93.
\bibitem{sirlin} A. Sirlin, J. Phys. {\bf G29}(2003)213;
P. Langacker, J. Phys. {\bf G29}(2003)1;
K. Matchev, TASI Lectures on Precision Electroweak Physics,
hep-ph/0402031.
\bibitem{zfitter} D. Bardin, P. Christova, M. Jack, L.
Kalinovskaya, A. Olshevski, S. Riemann and T. Riemann,
Fortran program package ZFITTER v.6.30 and description:
ZFITTER v.6.21: A semi-analytical program for fermion pair
production in $e^{+}e^{-}$ annihilation, 
Comp. Phys. Commun. {\bf 133} (2001)229.
\bibitem{opal2} OPAL Collaboration, G. Abbiendi {\it et al.},
Eur. Phys. J. {\bf C33}(2004)173.
\bibitem{opal3}OPAL Collaboration, G. Abbiendi {\it et al.},
Phys. Lett. {\bf B516}(2001)1; 
\bibitem{opal4} OPAL Collaboration, G. Abbiendi {\it et al.},
Phys. Lett. {\bf B519}(2001)23.
\end{thebibliography}
\end{document}